\documentclass[12pt]{article}
\usepackage{latexsym, epsfig, graphics}
\newcommand{\be}{\begin{equation}}
\newcommand{\ee}{\end{equation}}
\newcommand{\bq}{\begin{eqnarray}}
\newcommand{\eq}{\end{eqnarray}}
\begin{document}
\begin{titlepage}
\today          \hfill 
\begin{center}
\hfill    LBNL-53594 \\
          \hfill    UCB-PTH-03/21 \\

\vskip .5in

{\large \bf Self Consistent Field Method for Planar $\phi^{3}$ Theory}
\footnote{This work was supported in part
 by the Director, Office of Science,
 Office of High Energy and Nuclear Physics, 
 of the U.S. Department of Energy under Contract 
DE-AC03-76SF00098, and in part by the National Science Foundation
under Grant PHY-00-98840}
\vskip .50in


\vskip .5in
Korkut Bardakci

{\em Department of Physics\\
University of California at Berkeley\\
   and\\
 Theoretical Physics Group\\
    Lawrence Berkeley National Laboratory\\
      University of California\\
    Berkeley, California 94720}
\end{center}

\vskip .5in

\begin{abstract}
We continue and extend earlier work on the summation of planar
graphs in $\phi^{3}$ field theory, based on a local action
on the world sheet. The present work employs a somewhat different
version of the self consistent field (meanfield) approximation
compared to the previous work on the same subject. Using this new 
approach, we are able to determine in general the asymptotic forms
of the solutions, and in the case of one solution, even its exact form.
This solution leads to formation of an unstable string, in agreement
with the previous work. We also investigate and clarify questions
related to Lorentz invariance and the renormalization of the solution.

\end{abstract}
\end{titlepage}

\newpage
\renewcommand{\thepage}{\arabic{page}}
\setcounter{page}{1}
\noindent{\bf 1. Introduction}

\vskip 9pt

The summation of the planar graphs of a given field theory is an
old problem, going back to the pioneering contributions of Nielsen,
Olesen, Sakita and Virasoro [1, 2] and of 't Hooft [3]. This problem
has recently got a boost from the discovery of AdS/CFT correspondence
[4, 5].
 The general idea
is to construct a world sheet description of the planar graphs, which
will hopefully serve as the starting point of a dual string formulation
of at least some interesting field theories. Of these earlier work, the
most systematic one is due to 't Hooft [3]: He succeeded in constructing
a world sheet description of a $\phi^{3}$ field theory through the use
of mixed coordinate-momentum space light cone coordinates. This description
has, however, the drawback of being non-local, which limits its
usefulness.

By introducing additional non-dynamical fields, the original model
of 't Hooft was reformulated as a local world sheet theory in
reference [6]. The world sheet structure of a given graph consists
of the bulk and some solid lines, representing boundaries.
 All the dynamics is in the boundaries; the bulk merely transmits
the interaction between the boundaries. Although originally only
a $\phi^{3}$ theory was considered for simplicity [6], this
approach was later extended to more interesting and realistic
cases [7, 8].

 This  new local field theory on the world sheet has no new
physical content. Perturbatively, it will reproduce the usual
Feynman graphs. However, it has one great advantage:
 It is particularly well suited for the
investigation of the possibility of a
 dynamical phenomenon which we call condensation
of solid lines (boundaries). If this phenomenon takes place,
the dominant contribution to the sum over graphs comes from
graphs of asymptotically infinite order, with the solid
lines becoming dense on the world sheet. In this limit, one
can envisage a finite density of solid lines on the world sheet,
and the world sheet, which was originally purely topological,
becomes a dynamical object. It is then quite plausible that
condensation of the boundaries can provide the  mechanism
for string formation. The approximate calculations carried out in
section 7 support this connection between boundary condensation
and string formation.

 The possibility of boundary condensation
was investigated in references [9, 10]
 for the $\phi^{3}$ theory, using the mean field
or the self consistent field approximation. These two terms are
commonly used interchangeably; however, in this article we
prefer to use the second terminology, since our approach
stresses the feature of self consistency of the approximation.
 Although $\phi^{3}$
is an unphysical theory, it provides a simple setting for the
methods needed to attack more physical, but also more
complicated theories. The self consistent field approximation
is a widely used approximation scheme, well adapted to
the investigation of condensate formation. Judging from past
experience, there is reason to hope that it will at least lead
to qualitatively reasonable answers.

At this point, it is perhaps worthwhile to explain  what is
the justification for the present work and in what way it
 is different from reference [10]. It turns out that 
there is no unique way to do the self consistent field approximation:
The set up of the problem and the choice of fields can lead to
different approximation schemes. Although both are self consistent
field calculations, the scheme presented here is different from the
scheme used in [10]. The main difference lies in the choice of the
order parameters: Both in [10] and also in the present work,
 the order parameters are the ground state expectation values
of the product of two fields (two point functions).
However, in [10], both fields have the same $\sigma$ coordinate,
but different $\tau$ (time) coordinates on the world sheet, whereas
here, the fields have the same $\tau$ coordinate, but different
$\sigma$ coordinates. In one case the order parameters are
no-local in $\tau$; in the other case, they are non-local in $\sigma$.
 We will discuss this difference in more detail in section 5.
Since it is difficult to judge a priori the accuracy of any of these
approximation schemes, it seems reasonable to try as many different
approaches as possible and compare the results. Also, the present 
approach has a number of advantages over reference [10], some of which
we will discuss below. It is remarkable that despite the complicated
appearance of the self consistency equations, it is possible to find 
one exact solution, and determine the asymptotic form of the other
solutions.
 In addition, we give here a more complete discussion 
of the questions of cutoff dependence and renormalization, and we
investigate the problem of Lorentz invariance, which was not addressed
in [10] at all.

After a brief review,
we start with a local action 
for the field theory on the world sheet. The world sheet is taken to
be a continuum, and  fermions are introduced to keep 
track of the solid lines. This action is not really new; it is only
a slightly different version of one of the actions presented in
[6], and used in [9].
We believe that the introduction of a local action which
automatically keeps track of the solid lines greatly simplifies
and clarifies the subsequent approximations, which is one of the
advantages of the present treatment.

 In section 4, we investigate the question of the Lorentz 
invariance, which is obscured by the light cone formulation. We are able
to consider only a subgroup of the Lorentz group, under which
the fields have simple transformation properties. Among the generators
of this subgroup, invariance under one particular generator translates
into scale invariance for the action. This turns out to be the crucial
 restriction, which is easily violated by the introduction
of cutoffs later on. We also clarify the role of the the prefactor
$1/2 p^{+}$ in front of the propagator (eq.(2)), which turns out
to be the measure needed
to make the $p^{+}$ integrations scale invariant. This factor is
rather awkward to take care of directly in a continuum treatment
of the world sheet: It is easier to take care of it indirectly
by making sure that scale invariance is not violated.

The action of section 3 contains non-dynamical fields, and before
undertaking a self consistent field calculation, it is best to
eliminate them in favor of dynamical fields. This is done in
section 5. After this
elimination, the action becomes non-local in the $\sigma$ coordinate,
and it is also highly non-linear.  This action is the starting point
of the self consistent field approximation. The first step of this
approximation is to replace
  the exact action by a an action quadratic in
fields. This is done by a suitable cluster decomposition of the products
of fields in terms of their ground state irreducible components. In
the resulting expression, the coefficients of the quadratic terms in the
fields are expressible in terms of
  products of equal time two point functions. With the help of this action,
in section 6, we formulate the conditions for self consistency:
The two point functions, derived from the quadratic action, must
agree with the two point functions that are already present in the
same action as coefficients. Although the self consistency
conditions are complicated non-linear relations, it is possible to
push the calculations surprisingly far. In fact, if certain
conditions are satisfied, we have an exact result (eq.(45)) for
the two point function for the fermions. Also, some of the integrations
can be done explicitly, and it turns out that a cutoff is needed to
regulate a divergent integral. We show that, in order to
preserve Lorentz invariance, this cutoff must be redefined to include
a factor of $p^{+}$. We also show how to renormalize the model by
absorbing the cutoff into the definition of the renormalized coupling
constant.

In section 7, we develop asymptotic solutions to self consistency
equations in the large momentum, or equivalently, short distance
regime. In fact, when the fermion propagator is given by eq.(45),
it is possible to work out the exact solution. Since the resulting
formulas are somewhat cumbersome, we present only the asymptotic
form of the solution, which shows formation of a string, which
is however unstable. This result is in qualitative agreement
\footnote{ In a recent article [11], there was no string formation
when  the so called fishnet
graphs in the $\phi^{3}$ model
 were summed using mean field techniques with two order
parameters. However, since we are probably summing a different set of graphs,
it is not clear that there is a real disagreement.}
 with
with the earlier results [9,10], which employed technically
different versions of the same approximation scheme. There are
other possible solutions, but they are all even more unstable and
not particularly interesting.

The present work, in conjuction with the earlier work,
supports the hope that string formation is a generic phenomenon
in theories with attractive forces, such as the $\phi^{3}$
model considered here. Unfortunately, in this case, an unstable
field theory generates an unstable string. The expectation is
that, non-abelian gauge theories share the nice feature of string
formation with $\phi^{3}$, while avoiding the instability.

\vskip 9pt

\noindent{\bf 2. A Brief Review}
\vskip 9pt

We are interested in the sum over the planar graphs of zero mass
$\phi^{3}$ field theory.
In the mixed light cone representation of 't Hooft [3], the evolution
parameter (time) is $x^{+}$, and the conjugate Hamiltonian is
$p^{-}$, and the Euclidean evolution operator is given by 
\be
T= \exp\left(- x^{+} p^{-}\right).
\ee
 In this notation, a Minkowski vector $v^{\mu}$ has the light cone
components $(v^{+},v^{-},{\bf v})$, where $v^{\pm}=(v^{0}\pm v^{1})/
\sqrt{2}$, and the boldface letters label the transverse directions.
A Feynman propagator that carries momentum $p$
 is pictured as a horizontal strip of with $p^{+}$ and length
$\tau=x^{+}$ on the world sheet,
 bounded by two solid lines (Fig.1).
\begin{figure} [t]
\centerline{\epsfig{file=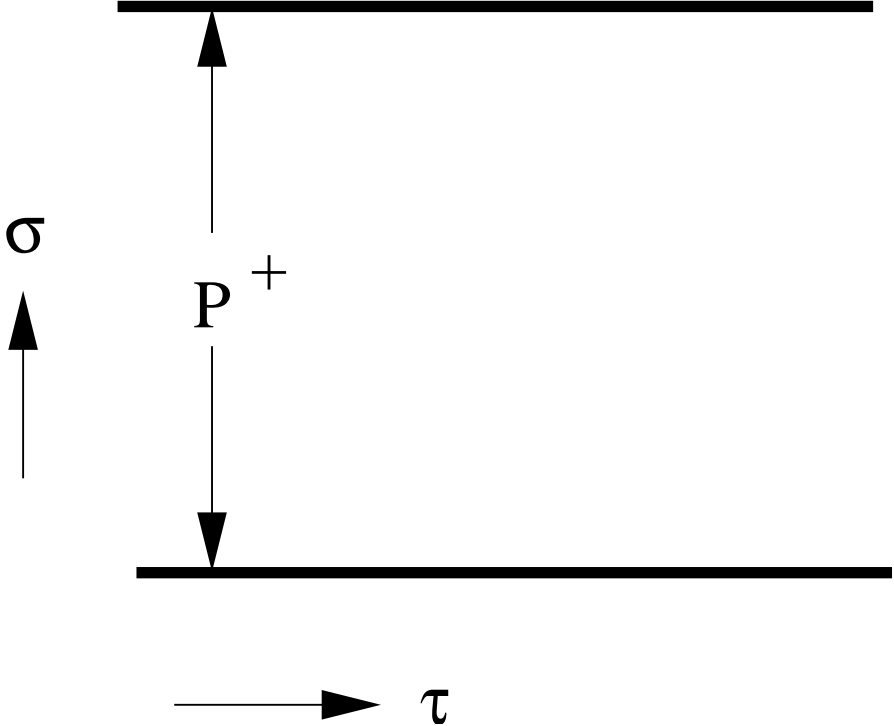, width=12 cm}}
\caption{The Propagator}
\end{figure}
 The solid lines
forming the boundary carry transverse momenta ${\bf q}_{1}$ and
${\bf q}_{2}$, where
$$
{\bf p}={\bf q}_{1}-{\bf q}_{2}.
$$
The corresponding Euclidean metric propagator is given by
\be
\Delta(p)=\frac{\theta(\tau)}{2 p^{+}}\exp\left(-\frac{\tau
({\bf q}_{1}-{\bf q}_{2})^{2}}{2 p^{+}}\right).
\ee

A more complicated graph, with interaction vertices, is shown
in Fig.(2). Interaction takes place at points where a solid line
begins or ends, and a factor of $g$ is associated with each such
point, where $g$ is the coupling constant. Finally, one has to
integrate over the position of the interaction vertices, as well
as the momenta carried by the solid lines.

It was shown in [6] that these Feynman rules can be reproduced 
by a local field theory defined on the world sheet. Parametrizing
the world sheet by the coordinate $\sigma$ in the $p^{+}$
direction and by $\tau$ in the $x^{+}$ direction,
the transverse momentum ${\bf q}$ is promoted 
 to a bosonic field
${\bf q}(\sigma,\tau)$ defined everywhere on the world sheet.
In addition, we introduce two fermionic fields (ghosts)
$b(\sigma,\tau)$ and $c(\sigma,\tau)$. In contrast to ${\bf q}$,
which has D components, $b$ and $C$ each have $D/2$ components.
Here, $D$ is the dimension of the transverse space, which is assumed to 
be even. The free part of the action on the world sheet is then
given by
\be
S_{0}= \int_{0}^{p^{+}}\!
d\sigma \int_{\tau_{i}}^{\tau_{f}} d\tau
\left(b'\cdot c' -\frac{1}{2}{\bf q}'^{2}\right),
\ee
where the prime denotes derivative with respect to $\sigma$. This
action is to be supplemented by the following Dirichlet boundary
condition on the solid lines:
\be
\dot{{\bf q}}=0,
\ee
where the dot denotes derivative with respect to $\tau$. Also, if the
transverse momentum carried by the whole graph is ${\bf p}$, the
constraint
\be
{\bf p}=\int_{0}^{p^{+}}\! d\sigma\,{\bf q}'
\ee
has to be imposed. In the rest of the paper, for simplicity, we will
set ${\bf p}=0$,
 which implies periodic boundary conditions on ${\bf q}$:
$$
{\bf q}(\sigma=0)={\bf q}(\sigma=p^{+}).
$$
We note that for timelike $p$, this can always be achieved by a
Lorentz transformation.

The boundary conditions on the ghosts corresponding to eq.(4) are
\be
b=c=0
\ee
on the solid lines.

The action $S_{0}$ plus the boundary conditions (4,6) reproduce only
the exponential factor in the expression for the propagator (eq.(2));
the factor in front, $1/2 p^{+}$, is missing. In [6], 
 this was taken care of by inserting a special vertex
built out of the ghost fields at the points of interaction. Since
this factor is intimately connected with Lorentz invariance, we
postpone its discussion until section 4.
\begin{figure}[t]
\centerline{\epsfig{file=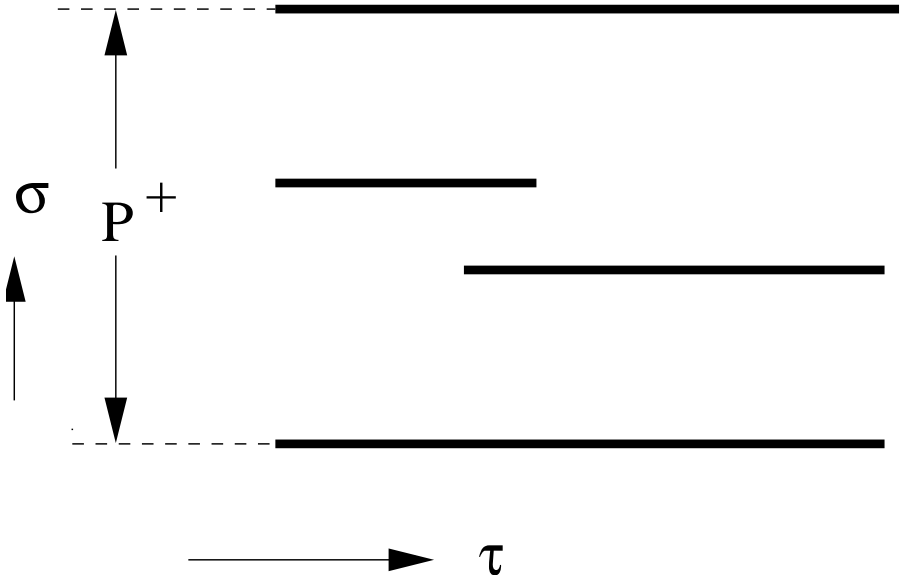, width=12cm}}
\caption{A Typical Graph}
\end{figure}
\vskip 9pt
\noindent{\bf 3. The Worldsheet Action}
\vskip 9pt

In this section, we are going to show how to
incorporate the boundary conditions
(4,6) into the action itself, rather than impose them by hand. This problem
was already addressed in [6], and here we will review especially the
treatment given in section 5 of this reference.
 The boundary
conditions will be implemented by introducing the Lagrange multipliers
${\bf y}(\sigma,\tau)$, $\bar{b}(\sigma,\tau)$ and $\bar{c}(\sigma,
\tau)$. For example, a solid line located at $\sigma=\sigma_{0}$
and running from $\tau=\tau_{1}$ to $\tau=\tau_{2}$ is associated 
with the term
\be
\Delta S= i \int_{\tau_{1}}^{\tau_{2}}\,d\tau\left({\bf y}(\sigma_{0},
\tau)\cdot \dot{{\bf q}}(\sigma_{0},\tau)+ \bar{b}(\sigma_{0},\tau)
\cdot b(\sigma_{0},\tau)+ \bar{c}(\sigma_{0},\tau)\cdot c(\sigma_{0},
\tau)\right),
\ee
in the action.
Carrying out the functional integral over the fields ${\bf y}$,
$\bar{b}$ and $\bar{c}$ results in the boundary conditions
$$
\dot{{\bf q}}(\sigma_{0},\tau)=0,\;\; b(\sigma_{0},\tau)=0,
\;\; c(\sigma_{0},\tau)=0,
$$
for $\tau_{1} \leq \tau \leq \tau_{2}$.

To reproduce the sum over all graphs, one has to associate the
beginning and the end of each solid line with a factor of $g$
(coupling constant), and then sum over all
non-overlapping solid lines and integrate over the location of each one.
In the above example of a single solid line, this amounts to a factor
of $g^{2}$ and to
integrating over $\sigma_{0}$ and $\tau_{1,2}$, subject to the
condition that $\tau_{1}< \tau_{2}$. In case of several solid lines
located at the same $\sigma$, there are obvious constraints on the
$\tau$ coordinates of the beginning and the end of each solid line
to avoid overlap between them.
 Rather then tackle this
problem directly, we will construct an action which automatically
takes care of the sum over the solid lines. For this purpose, we
 introduce a two component fermionic field $\psi_{i}(\sigma,\tau)$,
$i=1,2$, and its adjoint $\bar{\psi}_{i}(\sigma,\tau)$ on the world
sheet\footnote{These fields were denoted by $e_{i}$ and $\bar{e_{i}}$
in [6].}, and we  add to $S_{0}$ three new terms:
\be
S_{f}= S_{1} +S_{2} +S_{3},
\ee
where $S_{1}$ is given by
\be
S_{1}=\int_{0}^{p^{+}}\! d\sigma \int d\tau\:i \bar{\psi}\dot{\psi},
\ee
and it describes the free fermion propagation. The second term is
\be
S_{2}= -ig \int_{0}^{p^{+}}\! d\sigma \int d\tau\: \bar{\psi}
\sigma_{1}\psi,
\ee
 where $\sigma_{1}$ is one of the
Pauli matrices, acting on the two component fermion space, not to be
confused with the coordinate $\sigma$.\footnote{ When $\sigma$ carries a
numerical manuscript, it is a Pauli matrix; otherwise, it is the
sigma coordinate.}
 Finally, the third
 term is given by
\be
S_{3}=\int_{0}^{p^{+}}\! d\sigma\int d\tau\:
\left({\bf y}\cdot \dot{{\bf q}}+\bar{b}\cdot b+\bar{c}\cdot c\right)
 \bar{\psi}\left(\frac{1-\sigma_{3}}{2}\right) \psi.
\ee

 The contribution of the fermionic terms 
is easiest to figure out in the interaction representation. To switch
to this picture, we first quantize the free fermionic action $S_{1}$,
after  undoing the Euclidean rotation by letting
$g\rightarrow -ig$.  $S_{2}$ and $S_{3}$ are considered as
interaction terms and they are treated
perturbatively. The free action is quantized in terms of standard
 ladder operators with the anticommutation relations
$$
[\alpha_{i}(\sigma),\alpha_{j}^{\dagger}(\sigma')]_{+}=
\delta_{i,j}\, \delta(\sigma-\sigma').
$$
These operators satisfy the free equations of motion and consequently
they are $\tau$ independent. Now consider the Hamiltonian
corresponding to $S_{2}$,
\be
H_{2}= g \int_{0}^{p^{+}} d\sigma \left(\alpha^{\dagger}_{1}(\sigma)
\alpha_{2}(\sigma)+\alpha^{\dagger}_{2}(\sigma)\alpha_{1}(\sigma)\right).
\ee
This Hamiltonian, acting on the two states
\be
\alpha^{\dagger}_{i}(\sigma_{0})|0\rangle,\;\; i=1,2,
\ee
will change the state with $i=1$ into the state with $i=2$ and
vice versa. The state $|0 \rangle$, which we call the trivial
vacuum to distinguish it from the true ground state of the system,
is as usual defined by the condition
$$
\alpha_{1,2}(\sigma)|0\rangle = 0 .
$$
\begin{figure}[t]
\centerline{\epsfig{file=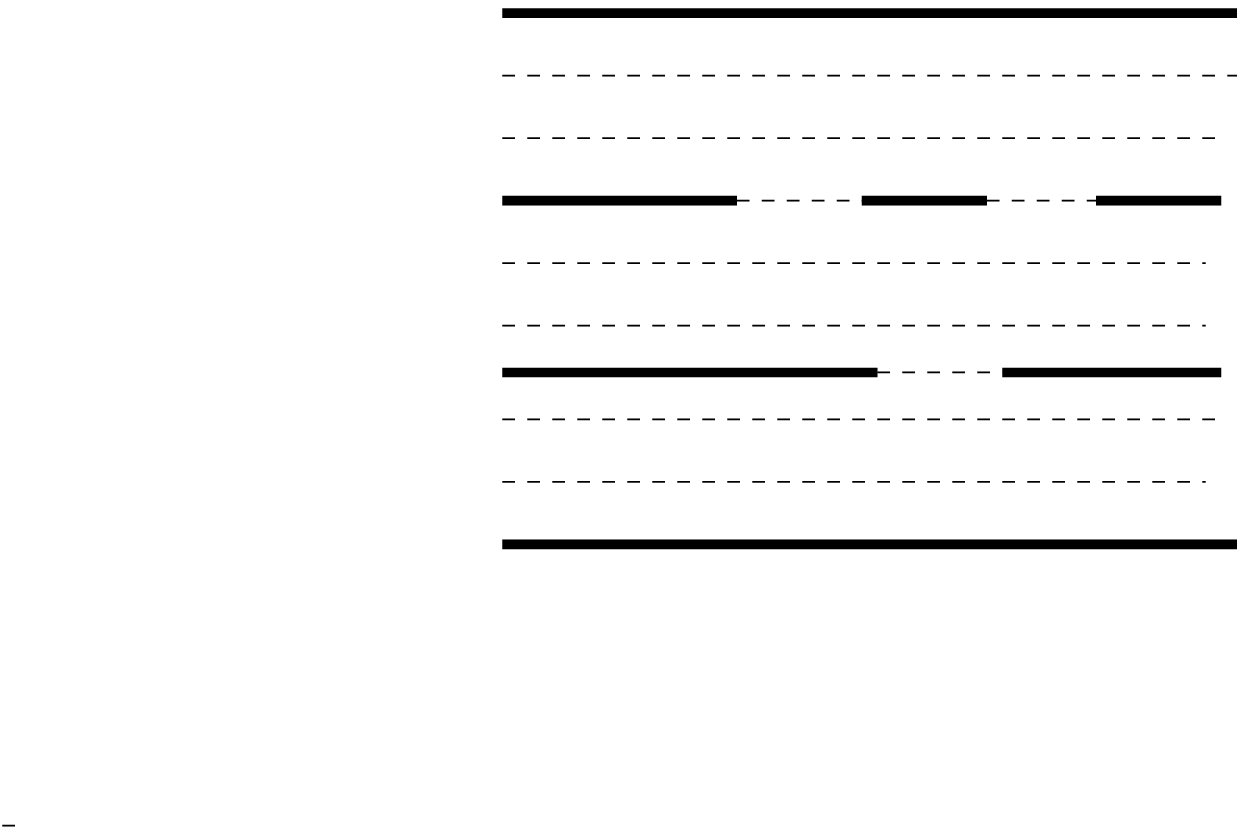, width=12cm}}
\caption{Solid and Dotted Lines}
\end{figure}

 We identify the state with $i=2$ as a solid line
located at $\sigma=\sigma_{0}$ and extending indefinitely for
both positive and negative $\tau$, and the state
with $i=1$ corresponds to the absence of a solid line at $\sigma=
\sigma_{0}$. For ease of exposition, we find it convenient 
to refer to the absence of a solid line at a given position by the
presence of a dotted line at the same position. Fig.(3) illustrates
the alternation between the solid and dotted lines corresponding
to the switch between the the states $i=2$ and $i=1$ under the 
action of $H_{2}$.

 However, this picture is not complete;
 in addition to the states given by (13), there are two
additional possible states: The trivial vacuum and the state
$$
\alpha^{\dagger}_{1}\alpha^{\dagger}_{2}|0\rangle.
$$
These states, which do not correspond to any graph on the
world sheet, can be eliminated
 by noticing that the unwanted states have fermion
numbers $0$ and $2$, as opposed to the ``good'' states (13), which
have fermion number $1$. Since fermion number is conserved, one
can consistently restrict the Hilbert space to the fermion
number $1$ sector by a suitable initial condition. We should
also like to point out the advantage of using fermions: If we
had used bosons, there would have been an infinite number of
unwanted states, which are absent in the case of fermions.

Now, let us examine $S_{3}$ given by eq.(11). We will quantize only
the fermionic fields and treat all others as classical external
fields. Since
$$
\bar{\psi}\left(\frac{1-\sigma_{3}}{2}\right)\psi=\alpha^{\dagger}_{2}
\alpha_{2},
$$
the Hamiltonian corresponding to $S_{3}$ gives zero acting on
a state with $i=1$ (dotted lines),
 and acting on a state with $i=2$ (solid lines), produces
$\Delta S$ given by (7). Recalling the discussion
following (7), we see that the correct boundary conditions on
the solid lines are enforced upon carrying out the functional
integrals over the fields ${\bf y}$, $\bar{b}$ and $\bar{c}$.
On the other hand, $\Delta S$  is missing in the case of the
dotted lines, and therefore there are no boundary conditions
associated with the dotted lines.
 We therefore see that the action $S_{f}$ given by eqs.(8,9,10,11) correctly
reproduces the dynamics of both the bulk and the boundaries.

For later applications, it will be more convenient to cast
 this action into a somewhat different form. So far,
${\bf y}$ was a non-dynamical Lagrange multiplier and ${\bf q}$
was the dynamical field.  The idea is to
interchange the roles of ${\bf y}$ and ${\bf q}$ by transferring
the $\tau$ derivative on ${\bf q}$ in eq.(11) to ${\bf y}$.
We do this in two steps: First by partial integration, we
transfer the $\tau$ derivative to ${\bf y}$ and the fermions.
Next, we use the equation 
\be
\partial_{\tau}\left(\bar{\psi}\frac{1-\sigma_{3}}{2}\psi\right)
+ig\, \bar{\psi}\sigma_{2}\psi=0,
\ee
which follows from the equations of motion for the fermions,
 to eliminate the $\tau$ derivative acting on them.
 As a result, we have
\bq
&&\int_{0}^{p^{+}}\!d\sigma\int d\tau\: {\bf y}\cdot\dot{{\bf q}}\:
\bar{\psi}\left(\frac{1-\sigma_{3}}{2}\right)\psi\nonumber\\
&&=\int_{0}^{p^{+}}\!d\sigma\int d\tau\:\bar{\psi}\left(ig\: {\bf y}
\cdot {\bf q}\: \sigma_{2} - \dot{{\bf y}}\cdot{\bf q}\:\frac{1-\sigma_{3}}
{2}\right)\psi.
\eq
Finally, putting everything together, we write down the complete action:
\bq
S&=&\int_{0}^{p^{+}}\! d\sigma\int d\tau\Big(b'\cdot c'-\frac{1}{2}
{\bf q}'^{2}+ \bar{\psi}\left(i\partial_{\tau} -ig \sigma_{1}\right)
\psi\nonumber\\
&+&\left(\bar{b}\cdot b+\bar{c}\cdot c -\dot{{\bf y}}
\cdot {\bf q}\right)\bar{\psi}\frac{1-\sigma_{3}}{2}\psi+ ig\: {\bf y}\cdot
{\bf q}\: \bar{\psi} \sigma_{2}\psi\Big).
\eq
 
In deriving this result, equations of motion were used to transform the
action. For the reader who is suspicious of this procedure, we sketch
an alternative derivation. Let us go back to $\Delta S$ (eq.(7)), which
took care of the boundary conditions on the solid line located at
$\sigma=\sigma_{0}$ and extending from $\tau=\tau_{1}$ to
$\tau=\tau_{2}$. Integrating by parts the first term on the right
gives
\be
i\int_{\tau_{1}}^{\tau_{2}} d\tau\, {\bf y}\cdot \dot{{\bf q}}
=- i \int_{\tau_{1}}^{\tau_{2}} d\tau\, \dot{{\bf y}}\cdot {\bf q}
+ i\left({\bf y}\cdot {\bf q}\right)_{\tau=\tau_{2}}
- i\left({\bf y}\cdot {\bf q}\right)_{\tau=\tau_{1}}.
\ee
We can now repeat the steps that lead to eq.(11). In the first term
on the right in that equation, the factor ${\bf y}\cdot \dot{
{\bf q}}$ will now be converted into $-\dot{{\bf y}}\cdot {\bf q}$.
There are in addition end point contributions, which occur when
a solid line turns into a dotted line or vice versa. Repeating the
arguments surrounding eq.(12), it is easy to see that the end point
contributions can be generated by an operator
\be
H'= i g\int_{0}^{p^{+}}\! d\sigma\left(\alpha^{\dagger}_{1}
(\sigma) \alpha_{2}(\sigma)- \alpha^{\dagger}_{2}(\sigma)
\alpha_{1}(\sigma)\right).
\ee
The minus sign between the two terms, as compared to the plus sign
in eq.(12), can be traced back to the difference in sign of the
end point contributions in (17). This Hamiltonian precisely corresponds
to an additional term in the action given by
$$
i g \int d\sigma \int d\tau \, {\bf y}\cdot {\bf q}\:
 \bar{\psi}\sigma_{2}\psi.
$$
Adding up various contributions, we get the same result as in
eq.(16).

We note that in eqs.(11,16), the ghost fields $b$ , $\bar{b}$, $c$,
and $\bar{c}$ have no $\tau$
(time) derivatives, which means that, in contrast to the usual
Faddeev-Popov ghosts, they are not propagating fields. This was
already observed in [9, 10], and in the Appendix of [10], it
was shown that ghosts
  do not change anything substantially
in the mean field or self consistent field
 computations. To simplify
matters, we will therefore drop the ghost terms :
\bq
S&\rightarrow& \int_{0}^{p^{+}}\! d\sigma \int d\tau\Big(
-\frac{1}{2} {\bf q}'^{2} + \bar{\psi}\left(i\partial_{\tau}
- i g \sigma_{1}\right) \psi -\dot{{\bf y}}\cdot {\bf q}
\, \bar{\psi}\left(\frac{1 -\sigma_{3}}{2}\right)\psi \nonumber\\
&+&i g\,{\bf y}\cdot {\bf q}\,\bar{\psi}\sigma_{2}\psi\Big).
\eq
This is the action that will be used in the rest of the paper.

Finally, we recall that ${\bf q}$ is a periodic function. As a
consequence, the equation
of motion with respect to ${\bf q}$ implies the following zero mode
condition:
\be
\int_{0}^{p^{+}}\!d\sigma\,\left(\bar{\psi}\left(\frac{1-\sigma_{3}}
{2}\right)\psi\,\dot{{\bf y}} -i g\bar{\psi} \sigma_{2}\psi {\bf y}
\right)=0.
\ee
This condition will be needed later on.

\vskip 9pt

\noindent{\bf 4. Lorentz Invariance}

\vskip 9pt

In this section, we will investigate the question of the Lorentz
invariance of the action (16). This is a non-trivial problem, since
the use of the light cone variables obscures the Lorentz transformation
properties of various dynamical variables. It is well known that
the light cone variables have simple transformation properties
under a special subgroup of the Lorentz group. If $L_{i,j}$ are the
angular momenta and $K_{i}$ are the boosts, this subgroup is
generated by the operators
\be
L_{i,j},\;\;M_{+,-}= K_{1},\;\;M_{+,i}= K_{i}+L_{1,i},
\ee
where the indices i and j run from 2 to D+2. Now let us consider
the light cone components $(v^{+},v^{-}, {\bf v})$ of a Minkowski
vector $v^{\mu}$. Under rotations in the transverse space generated
by $L_{i,j}$, $v^{+}$ and $v^{-}$ are invariant and ${\bf v}$
transforms like an ordinary vector.
 Under the finite transformations generated by $M_{+,-}$,
 the components transform as
\be
v^{+}\rightarrow a_{-1}\,v^{+},\;\;v^{-}\rightarrow a\,v^{-},
\;\;{\bf v}\rightarrow {\bf v},
\ee
where $a$ is the transformation parameter. Finally, the transformations
generated by $M_{+,i}$ lead to
\be
v^{+}\rightarrow v^{+},\;\; {\bf v}\rightarrow {\bf v}+ \sqrt{2}
v^{+} {\bf a},\;\;v^{-}\rightarrow v^{-}+\sqrt{2} {\bf a}\cdot
{\bf v}+2 v^{+} {\bf a}^{2},
\ee
where,again, ${\bf a}$ are the transformation parameters. Identifying
$v$ with $p$ or $x$ gives  the transformation rules for the
momentum and coordinate vectors.

Of the special transformations listed above, only rotations
in the transverse space and transformations generated by
$M_{+,-}$ are supposed to leave the action invariant. That is because
the action was originally derived from the evolution operator
given by eq.(1), and the combination $x^{+} p^{-}$ is invariant
under only these transformations. We will first consider this invariance
group of the action. The invariance of the matter part of the
action under rotations in the transverse space is obvious, since
${\bf q}$ and ${\bf y}$ transform as vectors. On the other hand,
the only way to ensure the invariance of the terms involving
the ghosts is to demand that under rotations, they behave like
scalars and do not transform at all. Next, let us consider
transformations under $M_{+,-}$.
 The field variables ${\bf q}(\sigma,\tau)$ and
${\bf y}$ themselves
are invariant, but  their arguments 
 $\sigma$ and $\tau$ are identified with
$p^{+}$ and $x^{+}$ respectively and must transform like them \footnote
{Actually, they must transform under the inverse transformations,
according to the well known field transformation rules.}. Therefore,
the Lorentz transformation generated by $M_{+,-}$ is identical to the
scale transformation:
\be
{\bf q}(\sigma,\tau)\rightarrow {\bf q}(a\sigma,a\tau),\;\;
{\bf y}(\sigma,\tau)\rightarrow {\bf y}(a\sigma,a\tau).
\ee
The ghost fields also scale in exactly the same way. Under scaling, 
the bulk of the action $S_{0}$ is then invariant, but the
boundary breaks scale invariance. In order to save Lorentz invariance,
the upper limit of integration in $\sigma$ must be changed by letting
\be
p^{+}\rightarrow p^{+}/a,
\ee
which is, after all, the correct transformation law for $p^{+}$
under $M_{+,-}$.

Now let us consider $S_{f}$, the fermionic part of the action.
 In order for $S_{1}$ (eq.(9)) to be invariant, we must
demand that
\be
\psi(\sigma,\tau)\rightarrow \sqrt{a}\, \psi(a\sigma,a\tau),\;\;
\bar{\psi}(\sigma,\tau)\rightarrow \sqrt{a}\, \bar{\psi}(a\sigma,
a\tau).
\ee
It then follows that $S_{3}$ is also invariant, but the interaction
term $S_{2}$ is not. The simplest way to fix it in the ghost free
action (19) is to redefine $g$ by letting
\be
g= g'/p^{+}.
\ee
Since $g$ is an effective constant defined on the worldsheet,
it is related to, but not identical to
 the fundamental constant of interaction of the original
field theory. If we identify $g'$ with the Lorentz and therefore
scale invariant coupling constant, there is no reason why
the relation between $g$ and $g'$ should not involve $p^{+}$.
In fact, Lorentz invariance determines the relation between
them uniquely to be eq.(27), and in the ghost free action, there
is no alternative to this identification.

Although we are not going to make use of the action (16) with ghosts,
it is of interest to note that in this case, there is an alternative
and perhaps a more fundamental way of restoring scale invariance.
We modify $S_{2}$ by introducing an extra factor involving ghosts:
\be
S_{2}\rightarrow -i g' \int_{0}^{p^{+}}\! d\sigma \int d\tau
\left(b\cdot c' + c\cdot b'\right) \bar{\psi} \sigma_{1}\psi,
\ee
and this factor supplies the missing scaling 
dimension. This may look like an ad hoc modification, but we recall that
 a similar factor was needed in reference [6] in the
latticized version of the worldsheet theory, and the factor we
have introduced could perhaps be identified with its continuum
counterpart.  We should
also  point out the close analogy with the open string
field theory [12], where a ghost insertion at the point of
interaction was necessary in order to preserve conformal invariance.
In the self consistent field approximation,
eqs.(27)and (28) are are actually connected. This approximation is equivalent
to replacing the factor with ghost fields by its ground state
expectation value:
$$
b\cdot c'+ c \cdot b' \rightarrow \langle b\cdot c' +
c\cdot b' \rangle,
$$
and using scaling arguments, it is easy to see that
$$
 \langle b\cdot c' +c\cdot b' \rangle = C/p^{+},
$$
where $C$ is a scaling invariant constant. 

 To understand
what is going on better, let us consider the propagator given
by (2). As part of a graph, this propagator will eventually be
integrated over $p^{+}$. As we have seen, the Lorentz transformation
generated by $M_{+,-}$ scales $p^{+}$; on the other hand, the
integration over $p^{+}$ by itself is not scale invariant, but
$$
\int \frac{d p^{+}}{p^{+}}
$$
is . It is then perhaps
more appropriate to think of the factor $1/2 p^{+}$
 as a scale invariant
 integration measure rather then as part of the propagator.
Integration measures are notoriously difficult to take care
of in field theory, since they are of quantum origin and strongly
sensitive to the regulator employed in computing quantum corrections.
If the integration measure is connected with invariance under some
symmetry, as is the case here, it is usually best to take care
of it indirectly by making sure that the symmetry is not violated.
 This is the approach we are taking here:
By demanding scale invariance, we believe that the integration
measure involving the $1/2 p^{+}$ factors is correctly fixed.
Of course, so far our treatment has been mainly classical; later
on, we shall see that quantum corrections can easily lead to
violations of scale invariance,
 which must be avoided to preserve Lorentz invariance.

Let us now briefly examine how the action transforms under the
remaining Lorentz generators $M_{+,i}$. For simplicity, we will
work with infinitesimal transformations. Although, as mentioned
earlier, the evolution operator (1) is not invariant under
$M_{+,i}$, it has simple transformation properties. In the exponent,
 the factor $x^{+}$ is
invariant, and $p^{-}$ transforms as (see (23))
\be
p^{-}\rightarrow p^{-}+ \sqrt{2}\, {\bf \epsilon}\cdot {\bf p},
\ee
where ${\bf \epsilon}$ is the vector infinitesimal parameter.
Now let us compare this with the transformation of the action
given by (19). Among the various terms, only ${\bf q}$  transforms:
\be
{\bf q}(\sigma,\tau)\rightarrow {\bf q}(\sigma,\tau) + \sqrt{2}
 \sigma {\bf \epsilon}.
\ee
In particular, ${\bf y}$ is invariant. Therefore, the only term in
the action that needs be considered is the first term on the right:
\bq
&&\int_{0}^{p^{+}}\! d\sigma \int d\tau\left(- \frac{1}{2}
{\bf q}'^{2}\right)\rightarrow \int_{0}^{p^{+}}\! d\sigma
\int d\tau\left(- \frac{1}{2} {\bf q}'^{2} -\sqrt{2} {\bf q}'
\cdot {\bf \epsilon}\right)\nonumber\\
&&= \int_{0}^{p^{+}}\! d\sigma \int d\tau \left( -\frac{1}{2}
{\bf q}'^{2}\right) - \sqrt{2} {\bf \epsilon}\cdot {\bf p}
\int d\tau,
\eq
where we have used eq.(5). Comparing with (29), we see that this is the
correct transformation law for the exponent of the evolution operator.
Since, in what follows, ${\bf q}$ will be eliminated from the
action, invariance under $M_{+,i}$ will be trivially satisfied.
In contrast, invariance under $M_{+,-}$, which translates into
scale invariance of the action, will be highly non-trivial.

This completes our discussion of Lorentz invariance. 
 There are, of
course, additional Lorentz generators which we have not discussed.
These are difficult to analyze, since their action results in the
change of the initial light cone frame. Although we will not pursue
this question in this article, demanding correct transformation of vectors
 under these generators leads to  non-linear relations. Such an
analysis was carried out in string theory, and it was shown that 
imposing Lorentz invariance in the light cone frame fixes the
critical dimension [13]. It would be interesting to see
whether this also happens when field theory is formulated on the
world sheet.

\vskip 9pt

\noindent{\bf 5. The Self Consistent Field Approximation}

\vskip 9pt

The action given by eq.(19), although simple looking, is not
easy to analyze. For one thing, both ${\bf q}$ and ${\bf y}$
are constrained fields and as such, they cannot easily be
quantized. We can at least partly overcome this problem by
explicitly carrying out the integration over ${\bf q}$. This gets
rid of a redundant field at the cost of introducing non-locality
in the $\sigma$ coordinate. The result is
\bq
S&=& \int_{0}^{p^{+}}\! d\sigma \int d\tau\,\bar{\psi}\left(
i\partial_{\tau} - ig \sigma_{1}\right)\psi\nonumber\\
&+& \frac{1}{4}\int_{0}^{p{+}}\!d\sigma \int_{0}^{p^{+}}\!
d\sigma' \int d\tau\, |\sigma-\sigma'|\Big(A_{1}\, \dot{{\bf y}}
_{\sigma}\cdot \dot{{\bf y}}_{\sigma'} + A_{2}\, \dot{{\bf y}}
_{\sigma}\cdot {\bf y}_{\sigma'}\nonumber\\
&+& A_{3}\,{\bf y}_{\sigma}
\cdot \dot{{\bf y}}_{\sigma'}+ A_{4}\,{\bf y}_{\sigma}\cdot
{\bf y}_{\sigma'}\Big),
\eq
where
\bq
A_{1}&=& - \frac{1}{4}\left(\bar{\psi}(1-\sigma_{3})\psi\right)_
{\sigma} \left(\bar{\psi}(1 -\sigma_{3})\psi\right)_{\sigma'},
\nonumber\\
A_{2}&=&\frac{i g}{2}\left(\bar{\psi}(1 -\sigma_{3})\psi\right)_
{\sigma} \left(\bar{\psi} \sigma_{2}\psi\right)_{\sigma'},
\nonumber\\
A_{3}&=& \frac{i g}{2}\left(\bar{\psi} \sigma_{2}\psi\right)_
{\sigma}\left(\bar{\psi}(1 -\sigma_{3})\psi\right)_{\sigma'},
\nonumber\\
A_{4}&=& g^{2}\left(\bar{\psi}\sigma_{2}\psi\right)_{\sigma}
\left(\bar{\psi}\sigma_{2}\psi\right)_{\sigma'}.
\eq
In the above equation, since all the fields have the same
$\tau$ coordinate, this coordinate
is suppressed , and the $\sigma$
coordinate is shown as a subscript. For example,
$$
\left(\bar{\psi}\sigma_{2}\psi\right)_{\sigma'}
\equiv \bar{\psi}(\sigma',\tau)\sigma_{2}\psi(\sigma',\tau).
$$
Also, the factor $|\sigma-\sigma'|$ is shorthand for the
 following Fourier series:
$$
\frac{1}{2} |\sigma -\sigma'|\equiv - \frac{1}{p^{+}} \sum_
{n\neq 0}
\left(\frac{p^{+}}{2\pi n}\right)^{2}
 \exp\left(\frac{2 \pi i n}{p^{+}}(\sigma -\sigma')\right),
\;\; n\in Z.
$$
The $n=0$ term, which would blow up, is absent because of 
the constraint (20).

Eqs.(32,33) for the action represents some kind of a generalized
string with coordinates ${\bf y}$, coupled to an Ising system
represented by the Fermionic fields $\psi$ and $\bar{\psi}$.
However, there is a complication: The string parameters $A_{n}$
are not numbers but composites in the fermionic fields. This
makes it difficult to disentangle the string sector from the
Ising sector. If however, the
$A_{n}$ develop non-zero expectation values about the ground
state; that is, if
$$
\langle A_{n}\rangle \neq 0,
$$
then, replacing $A_{n}$'s by their expectation values does result
in a familiar string action quadratic in the ${\bf y}$'s. This is,
of course, more complicated then the standard string action; for
example, it is in general non-local in the $\sigma$ coordinate. 
On the other hand, it is in the standard canonical form in the
time coordinate $\tau$, and it can easily be quantized. This
is the main difference between the present work and reference
[10]: In the treatment given there, the order parameters
 were local in $\sigma$ but
non-local in $\tau$. In contrast, we shall see below that
the order parameters of the present work are the two point
functions (eq.(35)) with the fields located at the same $\tau$,
but different $\sigma$ coordinates.

Non-zero expectation values for $A_{n}$'s means that some sort
of condensate has formed. Later, we will present evidence that
it is the solid lines that condense, which means that graphs
of very high (strictly speaking, infinite) order in the coupling
constant dominate. It is then quite possible that from the
sum of such graphs, a string picture emerges. The big challenge
is to decide this question by means of a dynamical calculation.
So far, the only dyanamical calculation that looks feasible is
based on the mean field approximation [9, 10, 11], which is
particularly well suited to the investigation of condensate
formation. This is an approximation scheme widely used in
many branches of physics. In the present application,
it can be derived by taking the large D limit, as was done
in these references . In the interests of brevity, we will
not reproduce this derivation, but instead, we will
describe it using the idea of self-consistent field or
the Hartree-Fock approximation.
 The basic idea is first to replace the terms in
 the action which are products of more then two fields
by terms quadratic in those fields, and then require self
consistency. One first systematically
 expands these products into a cluster
decomposition in terms of their vacuum irreducible components,
where by vacuum we mean the true ground state and not the trivial
vacuum of eq.(13).
 The approximation
consists of keeping only up to quadratic terms in fields
in this expansion and dropping
the rest. As an example, let us consider a typical term in eq.(32),
which has the general form
$$
T= \sum K_{aa',\alpha \alpha',\beta\beta'}\;
\left(y_{a} y_{a'} \bar{\psi}_
{\alpha}\psi_{\beta}\, \bar{\psi}_{\alpha'}\psi_{\beta'}\right),
$$
where various coordinates and indices the bosonic fields
$y$ and the fermionic fields $\psi$ and $\bar{\psi}$ carry
are replaced by a single index to simplify writing.
This term is approximated by
\bq
T&\rightarrow&\sum K_{a a',\alpha \alpha',\beta \beta'}
\Big(\Big[\langle \bar{\psi}_{\alpha}\psi_{\beta}\rangle
\bar{\psi}_{\alpha'}\psi_{\beta'} +\langle \bar{\psi}_{\alpha'}
\psi_{\beta'}\rangle \bar{\psi}_{\alpha}\psi_{\beta}\nonumber\\
&-& \langle \bar{\psi}_{\alpha} \psi_{\beta'}\rangle \bar{\psi}_
{\alpha'}\psi_{\beta}- \langle \bar{\psi}_{\alpha'} \psi_{\beta}
\rangle \bar{\psi}_{\alpha} \psi_{\beta'}\Big] \langle y_{a} y_{a'}
\rangle \nonumber\\
&+&\Big[\langle \bar{\psi}_{\alpha} \psi_{\beta}\rangle \langle
\bar{\psi}_{\alpha'}\psi_{\beta'}\rangle
-\langle \bar{\psi}_{\alpha}\psi_{\beta'}\rangle
\langle\bar{\psi}_{\alpha'}\psi_{\beta}\rangle\Big]
y_{a} y_{a'}\Big),
\eq
where  we have dropped pure c-number terms which will not be needed
in what follows.
To apply this to the action, we shall need the two point functions
(n=1,2,3)
\bq
\langle \partial_{\tau}^{(3-n)}{\bf y}(\sigma,\tau)
\cdot {\bf y}(\sigma',\tau)\rangle &=& G^{(n)}(\sigma -\sigma'),
\nonumber\\
\langle \bar{\psi}_{i}(\sigma,\tau) \psi_{i'}(\sigma',\tau)\rangle
&=& F_{i, i'}(\sigma -\sigma').
\eq
Because of the translation invariance of the ground state
 in both $\sigma$ and $\tau$
coordinates, these functions depend only on the difference
$\sigma-\sigma'$.  The approximate form of the action is
then given by
\bq
S&\rightarrow& \int_{0}^{p^{+}}\!d\sigma\int d\tau \bar{\psi}
\left(i\partial_{\tau} -i g \sigma_{1}\right)\psi\nonumber\\
&+& \frac{1}{4}\int_{0}^{p^{+}}\! d\sigma \int_{0}^{p^{+}}
\!d\sigma' \int d\tau\, |\sigma -\sigma'|\Big(a_{1}(\sigma -
\sigma')\, \dot{{\bf y}}_{\sigma}\cdot \dot{{\bf y}}_{\sigma'}
+a_{2}(\sigma-\sigma')\, \dot{{\bf y}}_{\sigma}\cdot {\bf y}_
{\sigma'}\nonumber\\
&+& a_{3}(\sigma -\sigma')\, {\bf y}_{\sigma}\cdot {\bf y}_{\sigma'}
+\sum_{i i'} b_{i, i'}(\sigma-\sigma')\: \bar{\psi}_{i,\sigma}
\psi_{i',\sigma'}\Big),
\eq
where,
\bq
a_{1}&=& F_{2,2}(\sigma-\sigma') F_{2,2}(\sigma'-\sigma)
- \left(F_{2,2}(0)\right)^{2},\nonumber\\
a_{2}&=& g F_{2,2}(\sigma-\sigma')\left(F_{2,1}(\sigma'
-\sigma)- F_{1,2}(\sigma' -\sigma)\right),\nonumber\\
a_{3}&=& g^{2}\Big(F_{1,2}(\sigma-\sigma') F_{1,2}
(\sigma'-\sigma)+ F_{2,1}(\sigma-\sigma') F_{2,1}(\sigma'
-\sigma)\nonumber\\
&-& 2 F_{1,1}(\sigma' -\sigma) F_{2,2}(\sigma-\sigma')
\Big),\nonumber\\
b_{1,1}&=& -2 g^{2} G^{(0)}(\sigma-\sigma') F_{2,2}(\sigma'
-\sigma),\nonumber\\
b_{1,2}&=& 2 g^{2} G^{(0)}(\sigma -\sigma') F_{1,2}(\sigma'
-\sigma) + 2 g\, G^{(1)}(\sigma -\sigma') F_{2,2}(\sigma'
-\sigma),\nonumber\\
b_{2,1}&=& 2 g^{2} G^{(0)}(\sigma -\sigma') F_{2,1}(\sigma'
-\sigma)+ 2 g\, G^{(1)}(\sigma -\sigma') F_{2,2}(\sigma'-\sigma), 
\nonumber\\
b_{2,2}&=& -2 g^{2} G^{(0)}(\sigma -\sigma') F_{1,1}(\sigma'
-\sigma) - 2\, G^{(2)}(\sigma -\sigma') F_{2,2}(\sigma' -\sigma)
\nonumber\\
&-& 2 g\, G^{(1)}(\sigma -\sigma') \left(F_{1,2}(\sigma' -\sigma)
+ F_{2,1}(\sigma' -\sigma)\right).
\eq
In arriving at this result, we took advantage of a number of simplifications:
For example, the symmetry of the integrand in the variables
$\sigma$ and $\sigma'$ was important. In addition, we have made use of
the relation
$$
F_{1,2}(0)= F_{2,1}(0),
$$
which follows by taking the vacuum expectation value of both sides of
 eq.(14). Also, an additive c-number term has been dropped. This term
contributes to the vacuum energy, but it is not needed for the self
consistency calculation that we are interested in.

The action (36) is quadratic in both the bosonic and fermionic fields;
it is therefore quite straightforward to compute the two point functions
given by (35) from it. Equating them to the same two point functions
appearing in (37) leads to the  self-consistency equations mentioned
earlier. These are non-linear equations for seven functions of a single
variable, namely:
$$
G^{(0)},\;G^{(1)}\;G^{(2)},\;F_{1,1},\; F_{1,2},\;
F_{2,1},\;F_{2,2}.
$$
In the next section, these equations will be developed further,
paying special attention to the cutoffs that are needed and
to Lorentz invariance.

\vskip 9pt

\noindent{\bf 6. The Self Consistency Conditions}

\vskip 9pt

We will first express the two point functions of eq.(35) in terms of 
$a_{n}$, $n=1,2,3$, and $b_{i,j}$ that appear in (36). We use the fact that
given a quadratic action of the bosonic field $X_{\alpha}$,
$$
S_{b}= \sum X_{\alpha} K_{\alpha \beta} X_{\beta},
$$
the two point function is
\be
\langle X_{\alpha} X_{\beta}\rangle = -\frac{1}{2} K^{-1}_{\alpha\beta},
\ee
and given a fermionic action
$$
S_{f}=\sum \bar{\psi}_{\alpha} K_{\alpha \beta} \psi_{\beta},
$$
the two point function is
\be
\langle \bar{\psi}_{\alpha} \psi_{\beta}\rangle = - K^{-1}_{\beta \alpha}.
\ee
Before applying this to the quadratic action (36) , it is best to go
to the momentum space. We define:
\bq
\frac{1}{2} |\sigma -\sigma'|\, a_{n}(\sigma -\sigma')&=&
 \frac{1}{p^{+}}\sum_{k}
r_{n}(k) \exp\left(i k (\sigma -\sigma')\right),\nonumber\\
\frac{1}{2} |\sigma -\sigma'|\, b_{i,j}(\sigma -\sigma')&=&
 \frac{1}{p^{+}}
\sum_{k}i\, t_{i,j}(k) \exp\left(i k (\sigma -\sigma')\right),
\nonumber\\
G^{(n)}(\sigma-\sigma')&=&\frac{1}{2\pi p^{+}} \sum_{k}
\exp\left(- i k (\sigma -\sigma')\right)\,\tilde{G}^{(n)}(k),
\nonumber\\
F_{i,i'}(\sigma -\sigma')&=&\frac{1}{2\pi p^{+}} \sum_{k}
\exp\left(- i k (\sigma -\sigma')\right) \tilde{F}_{i,i'}(k),
\eq
where,
$$
k=\frac{2\pi m}{p^{+}},\;\;m\in Z,
$$
because of the periodic boundary conditions.
From eqs.(38) and (39), it follows that
\bq
\tilde{G}^{(n)}(k)&=&-\int d k_{0}\, \frac{  (-i k_{0})^{3-n} D} 
 { k_{0}^{2}\, r_{1}(k)+ i k_{0}\,r_{2}(k) + r_{3}(k)},\nonumber\\
\tilde{F}_{i,i'}(k)&=&\int d k_{0} \left(\frac{2}{k_{0}+ i g \sigma_{1}
-t(k)}\right)_{i',i}.
\eq

The $k_{0}$ integrations are all elementary, and also, upon symmetrizing
the integrands with respect to the sign of $k_{0}$, they are all
finite except for the integral for $\tilde{G}^{(1)}$, which needs
a cutoff.
The results depend on the location of the poles of the integrand
in the complex $k_{0}$ plane. In the case of $\tilde{G}^{(n)}$'s,
 there are two poles corresponding to the zeros of the denominator,
and if
$$
r_{1} r_{3}<0,
$$
 then $Im(k_{0})$ has the same sign at both poles, and 
\bq
\tilde{G}^{(1)}(k)&=&\frac{D}{r_{1}(k)}\left(\Lambda -
\pi \frac{|r_{2}(k)|}{|r_{1}(k)|}\right),\nonumber\\
\tilde{G}^{(2)}(k)&=&\pi D \frac{|r_{2}(k)|}{r_{2}(k)|r_{1}(k)|},
\nonumber\\
\tilde{G}^{(3)}(k)&=& 0,
\eq
where $\Lambda$ is a suitable cutoff.
On the other hand, if
$$
r_{1} r_{3}>0,
$$
then $Im(k_{0})$ has opposite signs at the two
poles, and
\vskip 6pt
\bq
\tilde{G}^{(1)}(k)&=&\frac{ D \Lambda}{r_{1}(k)}\nonumber\\
&-&\pi D \left(\left(\frac{r_{2}(k)}{r_{1}(k)}\right)^{2}+ 2\frac{r_{3}(k)}
{r_{1}(k)}\right)\left((r_{2}(k))^{2}+ 4 r_{1}(k)r_{3}(k)\right)^
{-1/2} \frac{r_{1}(k)}{|r_{1}(k)|},\nonumber\\
\tilde{G}^{(2)}(k)&=&\pi D \frac{r_{2}(k)}{r_{1}(k)}\left(
(r_{2}(k))^{2} + 4 r_{1}(k) r_{3}(k)\right)^{-1/2}
\frac{r_{1}(k)}{|r_{1}(k)|}, \nonumber\\
\tilde{G}^{(3)}(k)&=& 2\pi D \left((r_{2}(k))^{2}+ 4 r_{1}(k) r_{3}(k)
\right)^{-1/2} \frac{r_{1}(k)}{|r_{1}(k)|}.
\eq
In writing down these results, we have tacitly assumed that all
$r(k)$'s are real, and that
$$
r_{2}^{2}+ 4 r_{1} r_{3} \geq 0.
$$
We shall see in the next section that, if these conditions are
not satisfied, we will have a complex valued action, which means
some kind of instability.

Next, consider the integral for $\tilde{F}_{i,i'}$. Upon symmetrizing
with respect to the sign of $k_{0}$, this integral is finite,
and the result again depends upon the location of the two poles.
They satisfy the quadratic equation
\bq
&&\det\left(k_{0}+ i g\sigma_{1} -i t(k)\right)\nonumber\\
&=&k_{0}^{2} - i k_{0}\left(t_{1,1}+ t_{2,2}\right)-
\left(t_{1,1} t_{2,2} - (t_{1,2}-g)(t_{2,1}-g)\right)=0.
\eq
To avoid again a complex valued action, we will  assume that
$t$ is a hermitian matrix. If the imaginary part of the poles
have the same sign, which happens if
$$
t_{1,1}t_{2,2}-(t_{1,2}- g)(t_{2,1}-g)> 0,
$$
then
\be
\tilde{F}_{i,i'}= 2\pi i\, \delta_{i,i'}\,\frac{t_{1,1}+ t_{2,2}}
{|t_{1,1}+ t_{2,2}|}.
\ee
If, on the other hand,
$$
t_{1,1} t_{2,2}- (t_{1,2} -g)(t_{2,1} -g)<0,
$$
then the poles have imaginary parts of opposite signs, and
\bq
\tilde{F}_{1,1}&=&\frac{2 \pi i}{R} \left(t_{1,1} -t_{2,2}\right),\nonumber\\
\tilde{F}_{1,2}&=&\frac{4 \pi i}{R} \left( t_{2,1} - g\right),\nonumber\\
\tilde{F}_{2,1}&=& \frac{4 \pi i}{R}\left(t_{1,2} - g\right),\nonumber\\
\tilde{F}_{2,2}&=& \frac{2 \pi i}{R} \left(t_{2,2} - t_{1,1}\right),
\eq
where,
$$
R=\left( (t_{1,1}- t_{2,2})^{2} + 4 (t_{1,2} -g)(t_{2,1} - g)\right)^{1/2}.
$$

We now have the complete set of equations needed to determine the
unknown functions. One could start with  $t_{i,i'}$
as the unknown functions to be determined by self consistency
 and go through the following chain of steps:\\
a) Express $\tilde{F}_{i,i'}$ in terms of $t_{i,i'}$ through
 through eq.(45) or eq.(46), and then Fourier transform to arrive
 at $F_{i,i'}$.\\
b) Use eq.(37) to determine $a_{1,2,3}$ and then use (40) to determine
the $r_{n}$, all in terms of $t_{i,i'}$.\\
c) Use the $r_{n}$ in eq.(42) or (43) to find $\tilde{G}^{(n)}$, and then
Fourier transform to find $G^{(n)}$.\\
d) Having already determined both $F_{i,i'}$ and $G^{(n)}$ in terms
of the t's go to
eq.(37) to find $b_{i,i'}$, and then use eq.(40) to find $t_{i,i'}$.\\
e) In this last step, the $t_{i,i'}$ are finally expressed in terms
of themselves. These are then the self consistency equations
for the $t_{i,i'}$.

As can be seen, this is a fairly complicated procedure. Most of the
steps are algebraic and easy to carry out, but there are also a
couple of Fourier transforms that cannot, in general, be done 
explicitly. However, if we start with the
 guess that the roots of eq.(44) have imaginary
parts of the same sign, there is one important simplification right
at the beginning. In this case, $\tilde{F}_{i,i'}$ are completely
determined up to an overall
 sign through eq.(45). It is then possible to write down
quite explicit formulas for all the other functions that appear in the
problem. The only self consistency check is to make sure at the end that
the imaginary part of the roots do indeed have the same sign. On the
other hand, if we start with the other possibility; namely, roots
with imaginary parts of opposite signs, it is not possible to
write down an explicit solution. In the next section, we will show
that, even in this case, it is possible to search for asymptotic
solutions valid for large momenta or short distances.

We would like to remark on the appearance of the cutoff $\Lambda$
in the expression for $\tilde{G}^{(1)}$. If we check invariance
under the scaling transformations (24) and (26), we find that the cutoff
violates scale invariance. This is not unexpected, since to start
with, scale invariance was only valid classically, and the cutoff
is associated with the quantum corrections. Although it would seem
that Lorentz invariance is lost, it is possible to save it by
redefining the cutoff:
\be
\Lambda=\frac{\Lambda'}{p^{+}}.
\ee
Just as in the case of the coupling constant $g$ (eq.(27)), the scaling
of $p^{+}$ (eq.(26)) restores Lorentz invariance. It is now easy to check
that with these redefinitions of the coupling constant and the cutoff,
the action does not depend on $p^{+}$ et all: The $p^{+}$ dependence
can be scaled out of the problem. This is clearly a consequence of 
Lorentz invariance.

It is also possible to scale away, at least partially, the dependence
on the coupling constant $g$. If in the action (19), we transform the
fields by
\bq
\psi(\sigma,\tau)\rightarrow \psi(\sigma, b\tau),&&
\bar{\psi}(\sigma,\tau)\rightarrow \bar{\psi}(\sigma,b\tau)\nonumber\\
{\bf q}(\sigma,\tau) \rightarrow \sqrt{b}\,{\bf q}(\sigma,b\tau),
&& {\bf y}(\sigma,\tau)\rightarrow \frac{1}{\sqrt{b}}\,
{\bf y}(\sigma, b\tau),
\eq
then $g$ scales by
\be
g \rightarrow g/b.
\ee
It might seem that the dependence on the coupling constant can be
eliminated by choosing $b=g$, but again, this is only a classical
result. Quantum mechanically, the cutoff enters the picture. From
the expression (42) or (43) for $\tilde{G}^{(1)}$, we see that the cutoff
must also transform:
\be
\Lambda\rightarrow \frac{\Lambda}{b}.
\ee
This means that the effective action does not depend on the coupling
constant and the cutoff seperately, but only on the combination
$$
\bar{g}=\frac{g}{\Lambda}.
$$
This  suggests the following renormalization process:
Define $\bar{g}$ to be the cutoff independent renormalized
coupling constant; then the theory is renormalized by noticing
that physical quantities like the effective action are expressible
in terms of only the renormalized coupling constant. This looks
nice, but the foregoing discussion of the cutoff dependence and
renormalization is really incomplete. This is because so far we
have dealt with only the divergence coming from the integration over
the variable $k_{0}$ in calculating the two point functions.
 In the next section, we will see
that there are divergences in various expressions coming from
the sums over discrete values of $k$, which makes it
necessary to introduce a second cutoff.

\vskip 9pt

\noindent{\bf 7. Asymptotic Solutions  }

\vskip 9pt

In this section, we will investigate the large momentum, or
equivalently, the short distance behaviour of the solutions
to the self consistency equations.
In addition to gaining some insight into the structure of the solutions,
we will be able to confront the issues of an additional cutoff and
renormalization. Let us start with eq.(45),
$$
\tilde{F}_{i,i'}= \pm 2\pi i\, \delta_{i,i'}
$$
which is one of the two options for $\tilde{F}$. We will consider
both signs; in fact, for most of our discussion, the sign will not
matter. As was pointed out earlier, it is now possible to compute
everything exactly. However, the resulting formulas are unwieldy,
and it is better to simplify matters as follows:
 Since we are only interested in the large $k$ asymptotic limit,
 the sum over the discrete values of $k$ can be replaced
 by an integral:
$$
\sum_{k=2\pi m/p^{+}}\rightarrow \frac{p^{+}}{2 \pi} \int d k.
$$
With this approximation, $F$ becomes a delta function:
$$
F_{i,i'}\rightarrow \pm i \delta_{i,i'} \delta(\sigma -\sigma').
$$
If we now go to eq.(37) for the $a$'s, we will encounter the square
of a  delta function, which is ill defined and needs regularization.
For simplicity, we will use a sharp cutoff $\lambda$
 as regulator, so that
\be
\delta(\sigma -\sigma')\rightarrow \delta_{\lambda}(\sigma -\sigma')
=\frac {1}{2\pi}\int_{-\lambda}^{\lambda} d k\, \exp\left(
i k (\sigma -\sigma')\right).
\ee
 Just as in the case 
of the previous cutoff, we have to set
\be
\lambda=\frac{\lambda'}{p^{+}},
\ee
in order to preserve Lorentz invariance.

It is now straightforward to compute the regularized versions of
the $a$'s and the $r$'s, with
\be
F_{i,i'}(\sigma)=\pm i \delta_{i,i'} \delta_{\lambda}(\sigma).
\ee
 We then need the following integrals:
\bq
&&\int_{0}^{p^{+}}\! d\sigma\, |\sigma|\,\exp(-i k \sigma)=
 -\frac{2}{k^{2}},\nonumber\\
&&\int_{0}^{p^{+}} \! d\sigma\, |\sigma|\,
 \delta^{2}_{\lambda}(\sigma)
\exp(- i k\sigma)\nonumber\\
&&= \frac{1}{(2\pi)^{2}}
\int \! d\sigma \int_{-\lambda}^{\lambda} d k_{1} \int_{-\lambda}^{\lambda}
 d k_{2}\, |\sigma|\,\exp\left( i \sigma (k_{1}+k_{2} - k)\right)\nonumber\\
&&= -2 \int_{-\frac{\lambda p^{+}}{2\pi}}^{\frac{\lambda
p^{+}}{2\pi}} d x_{1}\left(\int_{\frac{k p^{+}}{2\pi}+ 1 - x_{1}}^
{\frac{\lambda p^{+}}{2\pi}} + \int_{-\frac{\lambda p^{+}}{2\pi}}^
{\frac{k p^{+}}{2\pi} -1 - x_{1}}\right) dx_{2}\,\frac{1}{(k p^{+}
- 2\pi x_{1} - 2\pi x_{2})^{2}}\nonumber\\
&&=- \frac{p^{+}}{2\pi^{3}}(2\lambda - |k|) -\frac{1}{2\pi^{2}}
\ln\left(\frac{|k|}{2 \lambda(2\lambda -|k|)}\right).
\eq
The limits of integration over $k_{2}$ need an explanation. The factor
$$
\frac{1}{(k p^{+} - 2\pi x_{1} - 2\pi x_{2})^{2}}
$$
in the integrand
is singular at $k p^{+}=2\pi x_{1}+ 2\pi x_{2}$.
 However, this singularity is spurious:
It corresponds to the missing zero mode in the definition of $|\sigma|$ by 
Fourier series (following eq.(33)). When the discrete
Fourier sum is replaced by an integral, one should then cut out a suitable
 region  in the $(x_{1},\,x_{2})$ plane, as we
have done.

The regulated expressions for the $r$'s are given by
\bq
r_{1}(k)&=& \frac{1}{2} \int_{0}^{p^{+}}\! d\sigma\, |\sigma|\, \left(
F_{2,2}(\sigma) F_{2,2}(-\sigma) - \left(F_{2,2}(0)\right)^{2}
\right) \exp( -i k \sigma)\nonumber\\
&\rightarrow& -\frac{\lambda^{2}}{\pi^{2} k^{2}}
+\frac{p^{+}}{4 \pi^{3}}\left(2\lambda - |k|\right)+ \frac{1}{(2\pi)^{2}}
\ln\left(\frac{|k|}{2 \lambda (2\lambda - |k|)}
\right),\nonumber\\
r_{2}(k)&=&0,\nonumber\\
r_{3}(k)&=&-\frac{g^{2} p^{+}}{2 \pi^{3}}\left(2\lambda - |k|\right)-
 \frac{g^{2}}{2\pi^{2}}
\ln\left(\frac{|k|}{2 \lambda (2\lambda - |k|)}
\right),
\eq
where use has been made of the integrals in (54).

Let us try to see what sort of a string picture emerges from these
 results. Let us rewrite the ``stringy'' part of the action (36),
quadratic in ${\bf y}$, in momentum space. Setting
$$
{\bf y}(\sigma,\tau)=(2\pi p^{+})^{-1/2}\int d k_{0} \sum_{k}
\exp\left(i k \sigma + i k_{0} \tau\right) \tilde{y}(k, k_{0}),
$$
we have,
\be
S_{y}=\int d k_{0} \sum_{k}\left(k_{0}^{2}\,r_{1}(k)
- i k_{0}\,r_{2}(k) + r_{3}(k)\right) \tilde{y}(k, k_{0})
\cdot \tilde{y}(-k, -k_{0}).
\ee
With $r_{2}=0$, to have a well behaved Euclidean functional
integral, we must demand that
$$
r_{1}<0,\;\; r_{3}\leq 0.
$$
In the large $k$ and large cutoff regime, we can simplify (55)
by dropping the logarithmic terms:
\bq
r_{1}(k)&\simeq&-\frac{\lambda^{2}}{\pi^{2} k^{2}}+
\frac{p^{+}}{4\pi^{3}}\left(2\lambda -  |k|\right),\nonumber\\
r_{3}(k)&\simeq& -\frac{g^{2}p^{+}}{2\pi^{3}}\left(2 \lambda
- |k|\right),
\eq
and we see that, $r_{3}\leq 0$, for $|k|\leq 2\lambda$.
This last resriction is natural, since $\lambda$ is a sharp momentum
 cutoff. Although $r_{3}$ is well behaved, there is trouble
with $r_{1}$, since $r_{1}$ has a zero at a point
$$
k^{2}\approx \frac{2\pi \lambda}{p^{+}}.
$$
This means that the action has the right sign for 
$k^{2}$ smaller than this zero, but it has the wrong sign for 
the larger values of $k^{2}$, indicating instability. Of course,
such an instability is no surprise, since the field theory
we started with is unstable.

Despite the instability, it is of some interest to investigate
the string spectrum for the smaller values of $k^{2}$, when
there is stability. The spectrum can be read off 
 from the poles of the two point functions $\tilde{G}^{(n)}$:
\be
k_{0}^{2}\,r_{1}(k)+ i k_{0}\,r_{2}(k)+ r_{3}(k)=0.
\ee
Let us take
$$
k^{2}<< \frac{2\pi \lambda}{p^{+}}
$$
so that the first terms in the expression for $r_{1}$ and $r_{3}$
dominate:
\be
r_{1}(k)\approx -\frac{\lambda^{2}}{\pi^{2} k^{2}},\;\;
r_{3}(k)\approx -\frac{ g^{2}\lambda p^{+}}{\pi^{3}}.
\ee
Solving (58) for $k_{0}$ in this limit gives
\be
- k_{0}^{2}= \frac{r_{3}}{r_{1}}\simeq \frac{ g^{2} p^{+}}
{\pi \lambda} k^{2}.
\ee
Remembering that Euclidean $- k_{0}^{2}$ is Minkowski 
$+ k_{0}^{2}$, this spectrum gives rise to the usual 
string excitations that lie on linear trajectories. The
parameter
$$
\alpha^{2}=\frac{\lambda }{4 \pi g^{2} p^{+} },
$$
which is  the string slope parameter, should be finite.
We recall that in order to eliminate the dependence on the previous
cutoff, $g$ had to satisfy 
$$
g=\bar{g}\Lambda.
$$
 Expressing $\alpha$ in terms of the
renormalized coupling constant $\bar{g}$,
 it follows  that  the ratio
$$
\kappa=\frac{\Lambda^{2} p^{+}}{\lambda}
$$
should be finite.  Imposing this condition, we see that when
the finite parameters $\bar{g}$ and $\kappa$ are used, all the
cutoff dependence disappears.
The slope parameter $\alpha$ should also be Lorentz invariant,
or, when it is expressed in terms of $g'$ and
$\lambda'$, it should not depend on  $p^{+}$. This is easily
seen to be true. We also note that $\alpha\rightarrow \infty$
as $g\rightarrow 0$, which is physically reasonable.

To sum up, a nice consistent string picture emerges for relatively
small energies and momenta, before instabilities set in at higher
values of these quantities. In particular, a non-zero string
slope parameter crucially depended on having
$$
F_{2,2}(0)=\langle \bar{\psi}_{2}(\sigma,\tau) \psi_{2}(\sigma,\tau)
\rangle \neq 0.
$$
 In fact, the dominant term
for $r_{1}$ in eq.(59) is proportional to $(F_{2,2}(0))^{2}$.
From the discussion following eq.(12), we know that
$$
\bar{\psi}_{2}\psi_{2}
$$
represents the local number density for solid lines. A non-zero
expectation value for this operator is the signal for the
condensation of the solid lines. This then confirms the hypothesis
stated in the introduction: Condensation of the solid lines leads
to string formation, albeit an unstable one.

Unfortunately, because of the instability, we are unable to
verify the final step of self consistency; namely, whether the
zeros of eq.(44) are correctly located, so that eq.(45) is valid.
The problem is that, as a result of the instability, various
quantities like $r_{n}$ and $t_{i,i'}$ acquire imaginary parts,
contradicting the original assumption of reality.
 We can only hope that in a more
physical theory such as QCD, the string picture remains unchanged
but the instabilities and the problems caused by them are avoided.

We will now briefly discuss the alternative solution, where the
$F_{i,i'}$ are given by eq.(46). In this case, we can only get some
information about the asymptotic form of the solution for
large $k$. From scale invariance (eqs.(24,26))), which should be unbroken
for large $k$, we deduce that $\tilde{F}$'s, $t$'s and
 $r$'s tend to
$k$ independent constants as $k\rightarrow \infty$. When no confusion
can arise, we will denote these constants by the same letters:
\bq
\tilde{F}_{i,i'}(k)&\rightarrow& \tilde{F}_{i,i'},\nonumber\\
t_{i,i'}(k)&\rightarrow &t_{i,i'},\nonumber\\
r_{n}(k)&\rightarrow & r_{n}.
\eq
 One has then to recycle these constants through
the equations for self consistency in order to determine their
values. Without any hard work, it is easy to see that we can
consistently set
\be
t_{1,2}=t_{2,1}=g,\;\;r_{2}=0.
\ee
Since also, from eq.(46),
$$
\tilde{F}_{1,1}= -\tilde{F}_{2,2},
$$
following the steps that lead to eq.(55) for $r_{3}$, we find that
the sign of $r_{3}$ has changed; it is now positive. This because
$r_{3}$ is proportional to the product
$$
\tilde{F}_{1,1} \tilde{F}_{2,2},
$$
and this product has changed sign. A positive $r_{3}$ is again the signal
for instability. In this case, we do not even have the nice string
picture for smaller values of $k$. We have not investigated more
general possibilities which do not satisfy (62), but we suspect that
they are probably also unstable.

 The general features of the first solution (eq.(45)) are consistent
with the results obtained in the earlier work [9, 10]. There is string
formation at the lower end of the spectrum, but the string is unstable.
As we pointed out earlier, the approximation scheme used here is
technically somewhat different from those used in the earlier work.
 In view of this, the qualitative
agreement between them inspires some confidence in the methods used. 

\vskip 9pt

\noindent{\bf 8. Conclusions}

\vskip 9pt

In this article, we have reconsidered the problem of the summation
of the planar graphs in the $\phi^{3}$ theory, using a local action
formulation on the world sheet. This problem was already treated
in references [9, 10], with the help of mean field approximation.
In the present work, we start with a local action involving fermions,
 and we apply to it a version of the self consistent field
approximation, which is similar to but technically somewhat different
from the methods used in the earlier work. Although
the resulting equations look complicated, we are able to identify
one solution which, in principle, can be exactly calculated.
The high excitation asymptotic form of this solution indicates the
formation of an unstable string, in agreement with the earlier work
along the same lines. We also settle some questions related to
Lorentz invariance and the renormalization of the model.

The methods developed in this article and in related earlier work
 should be sufficently powerful to attack more physical
models, such as QCD. The world sheet formulation for such theories
is already available [7, 8], although some refinements are probably
needed to deal with the continuum limit. It remains to be seen
whether the self consistent field method is powerful enough to
give us useful information about string formation in these
theories.

\vskip 9pt

\noindent{\bf Acknowledgements}

\vskip 9pt

I would like to thank Charles Thorn for stimulating and valuable
discussions, especially concerning Lorentz invariance.
This work was supported in part
 by the Director, Office of Science,
 Office of High Energy and Nuclear Physics, 
 of the U.S. Department of Energy under Contract 
DE-AC03-76SF00098, and in part by the National Science Foundation
under Grant PHY-00-98840.

\vskip 9pt
 \noindent{\bf References}
\begin{enumerate}
\item H.B.Nielsen and P.Olesen, Phys. Lett. {\bf B 32} (1970) 203.
\item B.Sakita and M.A.Virasoro, Phys. Rev. Lett. {\bf 24} (1970)
1146.
\item G.'t Hooft, Nucl. Phys. {\bf B 72} (1974) 461.
\item J.M.Maldacena, Adv. Theor. Math. Phys. {\bf 2} (1998) 231,
[arXiv: hep-th/9711200].
\item O.Aharony, S.S.Gubser, J.M.Maldacena, H.Ooguri and Y.Oz,
Phys. Rept. {\bf 323} (2000) 183, [arXiv: hep-th/9905111].
\item K.Bardakci and C.B.Thorn, Nucl.Phys. {\bf B 626} (2002)
287, [arXiv: hep-th/0110301].
\item C.B.Thorn, Nucl. Phys. {B 637} (2002) 272, [arXiv:
 hep-th/0203167].
\item S.Gudmundsson, C.B.Thorn and T.A.Tran, Nucl. Phys.{\bf
B 649} (2003), [arXiv: hep-th/0209102].
\item K.Bardakci and C.B.Thorn, Nucl. Phys. {\bf B 652} (2003)
196, [arXiv: hep-th/0206205].
\item K.Bardakci and C.B.Thorn, Nucl. Phys. {\bf B 661} (2003)
235, [arXiv: hep-th/0212254].
\item C.B.Thorn and T.A.Tran, [arXiv: hep-th/0307203].
\item E.Witten, Nucl. Phys. {\bf B 268} (1986) 253.
\item P.Goddard, J.Goldstone, C.Rebbi and C.B.Thorn, Nucl. Phys.
{\bf B 56} (1973) 109.

\end{enumerate}

\end{document}